\documentclass[a4paper,11pt]{article}
\usepackage{etex}
\usepackage[utf8]{inputenc}
\usepackage{mathtools,amsmath,amssymb}
\usepackage{graphicx}
\usepackage{lmodern}
\usepackage[T1]{fontenc}
\usepackage{microtype}
\usepackage[pagebackref=false]{hyperref}
\numberwithin{equation}{section}

\usepackage{xspace}
\usepackage{bibentry}

\usepackage{graphicx}

\usepackage{subcaption}

\usepackage[numbers,sort&compress]{natbib}
\usepackage{hypernat}

\usepackage{tikz}
\usetikzlibrary{decorations.pathreplacing}
\usepackage[margin=2.3cm]{geometry}

\definecolor{todocolor}{HTML}{D7E1E5}
\usepackage[textsize=tiny, 
            backgroundcolor=todocolor, 
            bordercolor=todocolor, 
            linecolor=todocolor, 
            textwidth=3.9cm]{todonotes}

\usepackage{comment}
\usepackage{rotating}
\DeclareMathOperator{\diag}{diag}

\newcommand{\phil}{\beta_-}
\newcommand{\phir}{\beta_+}
\newcommand{\ma}{n}
\newcommand{\lws}{|\Phi_{2s+j}\rangle}
\newcommand{\psib}{\psi_{q}}
\newcommand{\psim}{ \widetilde\psi}
\newcommand{\gammab}{\Gamma_{q}}
\begin{document}

\begingroup\parindent0pt
% \begin{flushright}\footnotesize
% \texttt{preprint}\\
% % \texttt{HU-EP-17/14}\\
% % \texttt{TCDMATH-17-13}\\
% \end{flushright}
\vspace*{3em}
\centering
\begingroup\Large\bf
The non-compact XXZ spin chain as stochastic particle process
\par\endgroup
\vspace{3.5em}
\begingroup\large
{ Rouven Frassek}
\par\endgroup
\vspace{3em}
\begingroup
% \sffamily
% \footnotesize
Max-Planck-Institut für Mathematik, \\
   Vivatsgasse 7, 53111 Bonn, Germany\\
\par\endgroup
\vspace{5em}

% Version: \today

\vfill
\begin{abstract}
\noindent
In this note we relate the Hamiltonian of the integrable non-compact spin $s$ XXZ chain to the Markov generator of a stochastic particle process. The hopping rates of the continuous-time process are identified with the ones of a  q-Hahn asymmetric zero range model. The main difference with the asymmetric simple exclusion process (ASEP), which can be mapped to the ordinary XXZ spin chain, is that multiple particles can occupy one and the same site. For the non-compact spin $\frac{1}{2}$ XXZ chain the associated  stochastic process reduces to the multiparticle asymmetric diffusion model introduced by Sasamoto-Wadati. 
\end{abstract}
\endgroup

\vfill

\thispagestyle{empty}
\setcounter{tocdepth}{2}

% \newpage
\tableofcontents
\vfill
\newpage

\section{Introduction}
The XXZ spin chain remains one of the most studied integrable models solvable by Bethe ansatz \cite{Bethe1931}. Still it seems that there are many corners to be yet discovered and open problems to be solved, see e.g. \cite{baxter2016exactly,gaudin1983fonction,Faddeev:1996iy,korepin1997quantum} for some excellent literature on the topic. The bulk model is described by the nearest neighbor Hamiltonian 
\begin{equation}\label{eq:ham}
 H=\sum_{i=1}^{N-1}\mathcal{H}_{i,i+1}
\end{equation} 
which is given in terms of the Hamiltonian density $\mathcal{H}$ and the number of spin chain sites $N$.
Since the Leningrad school of the quantum inverse scattering method it is known that the Hamiltonian density of the Heisenberg spin chain admits a concise expression in terms of the digamma function which only depends on the representations in the irreducible tensor product decomposition of two neighboring sites \cite{Kulish:1981gi,Faddeev:1996iy}. This formula arises as the logarithmic derivative of the corresponding R-matrix which is written in terms of gamma functions.
The q-analogs of the Hamiltonian density and the R-matrix were given in \cite{Bytsko:2001uh}, see also  \cite{Chicherin:2014fqa,Mangazeev:2014gwa} for alternative presentations of the R-matrix.
In analogy to the rational case, the Hamiltonian density is written in terms of the q-analog of the digamma function and the R-matrix via the q-analog of the gamma function.  Despite the beauty of this expression it is not immediately obvious how the Hamiltonian density acts on the tensor product of two sites which is often what one would like to know when studying a physical problem. The change of basis can be fulfilled using Clebsch-Gordan coefficients which can however be quite cumbersome, especially when looking at non-compact spin chains. 

Stochastic particle processes of continuous time are usually defined through the master equation which is given in terms of the Markov matrix. The latter contains the rates with which the particles of the system move.  In the following we restrict ourselves to one-dimensional chains where particles can ``hop'' only to the neighboring sites. In this case the Markov matrix can be written in terms of local Markov generators as
\begin{equation}\label{eq:markov}
 M=\sum_{i=1}^{N-1}\mathcal{M}_{i,i+1}
\end{equation} 
similar to the Hamiltonian in \eqref{eq:ham}. 
It is well known that the Hamiltonian of the ordinary finite-dimensional XXZ chain can be  related to the Markov matrix of the asymmetric simple exclusion process (ASEP), see e.g. \cite{Mallick_2011,schutz2000exactly} for an overview. As a consequence the ASEP can be solved using Bethe ansatz methods. 
The Hamiltonian density of the ordinary finite-dimensional XXZ chain is usually written in terms of Pauli matrices at two neighboring sites. It belongs to the family of integrable XXZ chains mentioned above and can thus be written in terms of the digamma function, see \cite{Bytsko:2001uh}. One may expect that 
all XXZ spin chains that arise for different representations of $U_q(sl_2)$ from the integrable Hamiltonian density can be mapped to a stochastic particle system. This however does not seem to be the case. 
Furthermore, as the mapping depends on the choice of the basis, the stochastic process related to a Hamiltonian may be difficult to identify. 

Recently, a relation between rational non-compact spin chains and stochastic processes was pointed out in \cite{Frassek19}. The types of spin chains considered appeared previously in relation to high-energy QCD
and $\mathcal{N}=4$ super Yang-Mills theory
in the context of the AdS/CFT correspondence, see \cite{Beisert:2010jr} for further references as well as \cite{Derkachov:1999pz} and references thereof. 
It was found in \cite{Frassek19} that the Hamiltonian density of such spin chains yields the rational limits of a class of stochastic transport models that were known to be integrable. More precisely, the multiparticle asymmetric diffusion model (MADM) and a more general q-Hahn process that were defined in \cite{Sasamoto} and \cite{barraquand} respectively. Both arise from the zero range chipping model proposed in \cite{Povolotsky} which has been studied using commuting transfer matrices of the so-called stochastic R-matrix in \cite{Kuniba}.

Here we focus on connecting q-deformed non-compact (XXZ) spin chains with stochastic particle processes.  
% The representations studied in \cite{Bytsko:2001uh} were finite-dimensional, but  as the derivation of the R-matrix and Hamiltonian density is purely algebraic their form  immediately generalises to the infinite dimensional case. 
From the many aspects that were studied in \cite{Frassek19}, which presumably do have a q-analog, we restrict ourselves to setting the stage and relate the Hamiltonian of the non-compact XXZ chain to the Markov matrix of a stochastic q-Hahn process.

% Following \cite{Beisertt} we  

The paper is organised as follows. First we give some background about the relevant representations of $U_q(sl_2)$ and the non-compact XXZ spin chain, see Section~\ref{sec:bg}. In Section~\ref{sec:markov} we define the Markov process via the local Markov matrix which is given in terms of the rates of a q-Hahn process. Section~\ref{sec:toxxz} contains the relation between the Hamiltonian of non-compact XXZ chain and the Markov matrix of the stochastic process. This relation is proven, following \cite{Beisertt}, in the subsections. Further we study some limiting cases including the MADM and rational case in Section~\ref{sec:limit} and elaborate on an apparent connection between the TAZRP and Baxter Q-operators. Finally we conclude in Section~\ref{sec:conc}. In Appendix~\ref{app:qan} we collected the definition of some special functions.

\section{Hamiltonian density of the XXZ spin chain}\label{sec:bg}
The non-compact XXZ spin chain can be defined through its Hamiltonian density  acting on two neighboring sites of the spin chain 
\begin{equation}
 \mathcal{H}:D_s\otimes D_s\to D_s\otimes D_s\,.
\end{equation} 
Here $D_s$ denotes a spin $s$ lowest weight $U_q(sl_2)$ module. It is spanned by linearly independent vectors $|m\rangle$ with $m=0,1,\ldots$ on which the action of the q-deformed spin generators $S_\pm$ and $S_0$ is defined as
\begin{equation}\label{eq:sl2act}
 S_+|m\rangle=[m+2s]|m+1\rangle\,,\quad S_-|m\rangle=[m]|m-1\rangle\,,\quad S_0|m\rangle=(m+s)|m\rangle\,.
\end{equation} 
For simplicity we assume that the spin variable is a positive half integer, i.e.~$2s\in\mathbb{N}$.
The spin generators satisfy the standard $U_q(sl_2)$ commutation relations
\begin{equation}\label{eq:sl2}
 [S_+,S_-]=-[2S_0]\,,\quad\quad [S_0,S_\pm]=\pm S_\pm\,,
\end{equation} 
with the q-number
\begin{equation}
 [x]=\frac{q^{x}-q^{-x}}{q-q^{-1}}\,.
\end{equation}  
The Hamiltonian density is known to be invariant under $U_q(sl_2)$, i.e. it satisfies the commutation relations 
\begin{equation}\label{eq:comu}
 [\Delta(S_0),\mathcal{H}]=0 \,,\qquad [\Delta(S_\pm),\mathcal{H}]=0\,.
\end{equation} 
Here we have defined the co-product $\Delta$ which is given via 
\begin{equation}\label{eq:cop}
  \Delta(S_0)=S_0\otimes 1+1\otimes S_{0}\,,\qquad \Delta(S_\pm)=S_\pm\otimes q^{-S_0}+q^{S_0}\otimes S_{\pm}\,.
\end{equation} 
It can be shown to be compatible with  the commutation relations \eqref{eq:sl2}.
The Hamiltonian density can be derived purely algebraically, as done for the rational XXX spin chain in \cite{Kulish:1981gi,Faddeev:1996iy}, by solving the Yang-Baxter equation for certain R-operators. The logarithmic derivative then yields the Hamiltonian density. This has been done for the trigonometric XXZ chain in 
\cite{Bytsko:2001uh}. The result can be written as
\begin{equation}\label{ham:bytsko}
 \mathcal{H}(\mathbb{S})=\frac{\psib(\mathbb{S})-\psib(2 s)}{-q^{4 s} \log(q)}
\end{equation} 
with the q-analog of the digamma function $\psib(x)$ which we define following Bytsko as the logarithmic derivative $\psib(x)=\partial_x\log\gammab(x)$ of the q-analog of the Gamma function
\begin{equation}
 \gammab(x)=q^{\frac{1}{2}x(1-x)}(q^{-1}-q)^{1-x}\frac{(q^{2};q^{2})_\infty}{(q^{2x};q^{2})_\infty}\,,
\end{equation}  
for $|q|<1$ which we assume throughout the paper.\footnote{We note that the definition of the q-analogs deviate from the more standard conventions. The relation can be found in Appendix~\ref{app:qan}.} Further we introduced $(a;q)_\infty=\prod_{k=0}^\infty(1-aq^k)$.
The operator $\mathbb{S}$ in \eqref{ham:bytsko} is related to the two site Casimir via 
\begin{equation}
 \Delta(C)=[\mathbb{S}][\mathbb{S}-1]\,.
\end{equation} 
It can be computed explicitly using $
 C=[S_0][S_0-1]-S_+S_-$ and the action of the co-product defined in \eqref{eq:cop}. The action of the operator $\mathbb{S}$ is diagonal on each module in the irreducible tensor product decomposition 
\begin{equation}\label{eq:irrep}
 D_s\otimes D_s=\bigoplus_{j=0}^\infty D_{2s+j}\,,
\end{equation} 
and coincides with the action of $\Delta(S_0)$ on the corresponding lowest weight state.
\footnote{Note that we label non-compact representations with positive half-integers which would correspond to negative spin labels in \cite{Bytsko:2001uh}. The Hamiltonian density \eqref{ham:bytsko} can be related, up to a term proportional to the identity, to the one in \cite{Bytsko:2001uh} using the reflection property of $\psi_q$ as given in \eqref{psi:refl} and taking $\mathbb{S}\to-\mathbb{S}$.} More precisely, given a lowest weight state $\lws$ we have
\begin{equation}\label{eq:lws}
\Delta(S_-)\lws=0\,,\qquad\Delta(S_0)\lws=\mathbb{S}\lws=(2s+j)\lws\,.
\end{equation} 
As it has been discussed at various points in the literature, the form of  Hamiltonian density in \eqref{ham:bytsko} is not very convenient to study its action on the tensor product of two sites. In principle it can be obtained using Clebsch-Gordan coefficients, see e.g. \cite{koelink1998convolutions}.  The change of basis is however quite involved. We will come back to this problem later in Section~\ref{sec:toxxz}.

\section{Markov generator and the q-Hahn process}\label{sec:markov}
In the following we give the local Markov matrix that is obtained from the transfer rates of the q-Hahn process. These arise from the model introduced by Povolotsky \cite{Povolotsky} and were given in \cite{barraquand}.

Let $(m,m')$ denote a configuration at two neighboring sites with $m$ particles at the first and $m'$ particles at the second site. 
The local generator of the Markov process $\mathcal{M}$ describes the transition rates to a configuration $(\tilde m,\tilde m')$
\begin{equation}
 \mathcal{M}:(m,m')\mapsto(\tilde m,\tilde m')\,,
\end{equation} 
where the particle number $n$ is conserved, i.e. we have $\ma=m+m'=\tilde m+\tilde m'$.
For our purposes it is convenient to represent the Markov generator as an $(\ma+1)\times(\ma+1)$ matrix acting on the basis defined via the identification
\begin{equation}
 (n-i,i)\longleftrightarrow e_{i+1}^\ma\,.
\end{equation} 
Here $e_{i+1}^\ma$ denotes the basis vector of size $\ma+1$ such that $\left(e_{i}\right)_j=\delta_{ij}$.
Thus for a given number of particles $n$ the local Markov matrix is of the form
\renewcommand{\arraystretch}{1.4}
\begin{equation}\label{eq:locmarkov}
{\small
\mathcal{M}_\ma=\left(
 \begin{array}{ccccc}
  \alpha_+(\ma)+  \alpha_-(0)&-\phil(1,1)&-\phil(2,2)&\cdots&-\phil(\ma,\ma)\\
 -\phir(\ma,1)&  \alpha_+(\ma-1)+  \alpha_-(1)&-\phil(2,1)&\cdots&-\phil(\ma,\ma-1)\\
  -\phir(\ma,2)&-\phir(\ma-1,1)&\alpha_+(\ma-2)+  \alpha_-(2)&\cdots&-\phil(\ma,\ma-2)\\
  \vdots&\vdots&\vdots&\ddots&\vdots\\
  -\phir(\ma,\ma)&-\phir(\ma-1,\ma-1)&-\phir(\ma-2,\ma-2)&\cdots&  \alpha_+(0)+  \alpha_-(\ma)
 \end{array}
\right)\,.
}
\end{equation} 
Here $\beta_+(m,k)$ ($\beta_-(m,k)$) denotes the rate with which $k$ of $m$ particles  at the left (right) site are moved to the right (left) site. The diagonal entries are given in terms of the rates $\beta_\pm(m,k)$ via
\begin{equation}
\alpha_+(m)=\sum_{k=1}^{m} \beta_+(m,k)\,,\qquad 
 \alpha_-(m)=\sum_{k=1}^{m} \beta_-(m,k)\,.
\end{equation} 
This ensures that the sum over all columns vanishes for any $n$ which is a necessary requirement of the local Markov matrix.
The hopping rates for particles moving to the right take the form
\begin{equation}\label{eq:bp}
\beta_+(m,k)= 
\frac{\mu ^{k} }{\mu \left(1-\gamma^k\right)}\frac{(\gamma;\gamma)_m (\mu ;\gamma)_{m-k}}{ (\gamma;\gamma)_{m-k} (\mu ;\gamma)_m}\,,
\end{equation} 
and the rates for particles moving to the left are given by
\begin{equation}\label{eq:bm}
\beta_-(m,k)= 
\frac{1}{\mu  \left(1-\gamma^k\right)}\frac{(\gamma;\gamma)_m (\mu ;\gamma)_{m-k}}{ (\gamma;\gamma)_{m-k} (\mu ;\gamma)_m}\,.
\end{equation}  
The parameters take the values $0<\gamma,\mu<1$.
Further  we introduced the q-Pochhammer symbol  $(a;\gamma)_m=\prod_{j=0}^{m-1}(1-a\gamma^j)$.
The diagonal entries of the local Markov matrix \eqref{eq:locmarkov} can then be computed. We find
\begin{equation}
\alpha_+(m)
=\sum _{k=0}^{m-1} \frac{\gamma^k}{1-\mu  \gamma^k}\,,\qquad 
 \alpha_-(m)=\frac{1}{\mu}\sum _{k=0}^{m-1} \frac{1}{1-\mu  \gamma^k}\,.
\end{equation} 
The transition rates of the q-Hahn process were given in  \cite[where $\mu=\nu$ and $\gamma=q$]{barraquand}. 
We remark that the left and right hopping rates introduced in \cite{barraquand} are weighted by the parameters $R$ and $L$ that are fixed in our setup, $R=(1-\gamma)^{-1}$ and $L=\mu^{-1}(1-\gamma)^{-1}$. We discuss how these parameters can possibly be incorporated into our framework in the last paragraph of  Section~\ref{sec:limit}. 
Further we note that the process defined coincides with the $U_\gamma(A_1^{(1)})$ zero range process in \cite[equation (36) for $a=1$ and $b=\mu^{-1}$]{Kuniba:2016tzg} which was obtained from the stochastic R-matrix \cite{Kuniba}. 

In the next section we will show that the Markov process \eqref{eq:markov} defined via the local Markov matrix \eqref{eq:locmarkov} is directly related to the  Hamiltonian of the non-compact spin $s$ XXZ chain \eqref{eq:ham} defined via the Hamiltonian density \eqref{ham:bytsko}.

\section{From the q-Hahn process to the  XXZ spin chain}\label{sec:toxxz}

In order to connect the Markov generator of the q-Hahn process in \eqref{eq:markov} with the Hamiltonian of the XXZ chain in \eqref{eq:ham} we first identify
\begin{equation}\label{eq:iden}
 \gamma=q^2\,,\qquad \mu=q^{4s}\,.
\end{equation} 
Here $q$ denotes the deformation parameter and $s$ the spin label that were introduced in Section~\ref{sec:bg}. We first focus on the local Markov generator $\mathcal{M}_\ma$. For a given number of particles $n$ we then define the matrix 
\begin{equation}\label{eq:hammag}
 \mathcal{H}_\ma=\mathcal{S}_\ma\,\mathcal{M}_\ma\, \mathcal{S}_\ma^{-1}\,,
\end{equation} 
where the similarity transformation for fixed $n$ only depends on the spin variable. It reads
\begin{equation}
 \mathcal{S}_\ma=\diag\left(\begin{array}{cccccc}
                 1&q^{-2s}&q^{-4s}&\cdots&q^{-2\ma s}
                \end{array}
\right)\,.
\end{equation} 
As we will see, $\mathcal{H}_n$ is the Hamiltonian density of the integrable non-compact spin $s$ XXZ chain for a given magnon block of $n$ magnons.
In order to reveal the algebraic structure we write down how the Hamiltonian density acts on the tensor product of two $U_q(sl_2)$ modules as defined in \eqref{eq:sl2act}. We find that it can be written as
\begin{equation}\label{eq:haction}
\begin{split}
 \mathcal{H}|m\rangle\otimes|m'\rangle=\left(\alpha_+(m)+\alpha_-(m')\right)|m\rangle\otimes|m'\rangle&-\sum_{k=1}^{m}\rho(m,k)|m-k\rangle\otimes|m'+k\rangle\\
 &-\sum_{k=1}^{m'}\rho(m',k)|m+k\rangle\otimes|m'-k\rangle\,,
 \end{split}
\end{equation} 
with $m+m'=n$ and  the off-diagonal coefficients 
\begin{equation}\label{eq:rho}
\begin{split}
 \rho(m,k)&=\frac{q^{2ks}}{q^{4s} \left(1-q^{2k}\right)}\frac{(q^2;q^2)_m (q^{4s} ;q^2)_{m-k}}{ (q^2;q^2)_{m-k} (q^{4s} ;q^2)_m}\,.
 \end{split}
\end{equation} 
We note that the similarity transformation symmetrised the hopping coefficients \eqref{eq:bp} and \eqref{eq:bm} as $q^{-2ks}\beta_+(m,k)=q^{2ks}\beta_-(m,k)=\rho(m,k)$ under the identification \eqref{eq:iden}. As the diagonal terms remain unchanged by the similarity transformation, $\mathcal{H}_n$ is not a Markov matrix as the sum over its columns is non-vanishing. Using the variables $q$ and $s$ as given in \eqref{eq:iden} the terms on the diagonal can nicely be written in terms of the q-analog of the digamma function, cf.~Appendix~\ref{app:qan}, as
\begin{equation}\label{eq:alpha}
 \alpha_\pm(m)
 =\frac{\psib\left(m+2s\right)-\psib\left(2s\right)\pm m\log(q)}{-2q^{4s}   \log(q)}\,.
\end{equation} 
The Markov process \eqref{eq:markov} can then be related to the Hamiltonian \eqref{eq:ham} via
% \begin{equation}\label{sim:trans}
%  \mathcal{H}=(1\otimes q^{-2s(S_0-s)})\mathcal{M}(1\otimes q^{2s(S_0-s)})\,.
% \end{equation} 
\begin{equation}\label{sim:trans}
 M=q^{2s\sum_{k=1}^N k\, S_0^{[k]}}\,H\,q^{-2s\sum_{k=1}^N k\, S_0^{[k]}}\,.
\end{equation} 
This can be shown by acting on states using the action of the Hamiltonian density \eqref{eq:haction}. Here $S_0^{[k]}$ denotes the spin operator $S_0$ acting on site $k$. The identification \eqref{sim:trans} holds term by term on the level of the local Markov generator $\mathcal{M}_{i,i+1}$ and the Hamiltonian density $\mathcal{H}_{i,i+1}$. It follows that the closed Markov chain of length $N$ with $\mathcal{M}_{N,N+1}=\mathcal{M}_{N,1}$ can be related to a spin chain with diagonal twist such that $\mathcal{H}_{N,N+1}=q^{2sNS_0^{[1]}}\mathcal{H}_{N,1}q^{-2sNS_0^{[1]}}$. A similar transformation can be done for open spin chains with non-trivial boundaries in order to take into account the similarity transformation in the boundary terms at site ``1'' and ``$N$''.

In the remaining part of this section we show that the Hamiltonian density defined via \eqref{eq:hammag} or equivalently by the action \eqref{eq:haction} can indeed be identified with the  Hamiltonian density \eqref{ham:bytsko} of the non-compact XXZ spin chain. Following \cite{Beisertt}, we first show that $\mathcal{H}$ commutes with the co-product of the generators $S_\pm$ and $S_0$ 
\begin{equation}
 [\Delta(S_0),\mathcal{H}]|m\rangle\otimes|m'\rangle=0\,,\qquad  [\Delta(S_\pm),\mathcal{H}]|m\rangle\otimes|m'\rangle=0\,,
\end{equation} 
cf.~\eqref{eq:comu}. And as a second step we then show that the eigenvalues of the Hamiltonian density \eqref{eq:haction} coincide with the Hamiltonian density in \eqref{ham:bytsko} by acting on lowest weight states $\lws$ of the irreducible tensor product decomposition \eqref{eq:irrep}, i.e. 
\begin{equation}
 \mathcal{H}\lws=\frac{\psib(2s+j)-\psib(2 s)}{-q^{4 s} \log(q)}\lws\,.
\end{equation}

\subsection{$U_q(sl_2)$ invariance of the Hamiltonian density}
In this subsection we show that the Hamiltonian density defined by \eqref{eq:haction} is $U_q(sl_2)$ invariant or in other words commutes with the co-product of the generators as described in Section~\ref{sec:bg}.

The relation involving the Cartan generator $S_0$ is easily verified.  As the particle number is conserved in every magnon block we find
\begin{equation}
 [\Delta(S_0),\mathcal{H}]|m\rangle\otimes|m'\rangle=0\,.
\end{equation} 
Verifying the commutation relations involving the creation and annihilation operators $S_\pm$ is more involved, i.e. 
\begin{equation}
 [\Delta(S_\pm),\mathcal{H}]|m\rangle\otimes|m'\rangle=0\,.
\end{equation} 
To verify that it indeed holds, we explicitly evaluate the action of the commutator on the states using
\begin{equation}
\begin{split}
 \Delta(S_-)|m\rangle\otimes|m'\rangle
 &=[m]q^{-m'-s}|m-1\rangle\otimes|n\rangle+q^{m+s}[m']|m\rangle\otimes|m'-1\rangle
 \end{split}\,,
\end{equation} 
and
\begin{equation}
\begin{split}
 \Delta(S_+)|m\rangle\otimes|m'\rangle
 &=[m+2s]q^{-m'-s}|m+1\rangle\otimes|n\rangle+q^{m+s}[m'+2s]|m\rangle\otimes|m'+1\rangle\,,
 \end{split}
\end{equation} 
as well as the action of the Hamiltonian density given in \eqref{eq:haction}. 
For the annihilation operators $S_-$ we then find 
\begin{equation}
\begin{split}
  [\Delta(S_-),\mathcal{H}]|m\rangle\otimes|m'\rangle&=-\sum_{k=0}^{m-2}A_k^-(m,m')|k\rangle\otimes|m'+m-k-1\rangle+B^-(m,m')|m-1\rangle\otimes|m'\rangle\\
 &\quad\,-\sum_{k=0}^{m'-2}C_k^-(m,m')|m+m'-k-1\rangle\otimes|k\rangle+D^-(m,m')|m\rangle\otimes|m'-1\rangle
  \end{split}
\end{equation} 
where the coefficients are given by 

\begin{align} 
 A_k^-(m,m')&=
\rho(m,m-k-1)[k+1]q^{-m'-m+k+1-s}+\rho(m,m-k)q^{k+s}[m'+m-k]\\&\quad\, -[m]q^{-m'-s}\rho(m-1,m-k-1)-q^{m+s}[m']\rho(m,m-k)\nonumber\\[3mm]
B^-(m,m')&=[m]q^{-m'-s}\alpha_+(m)-\rho(m,1)q^{m-1+s}[m'+1]\\&\quad\,-
[m]q^{-m'-s}\alpha_+(m-1)+q^{m+s}[m']\rho(m,1)\nonumber\\[3mm]
C_k^-(m,m')&=\rho(m',m'-k)[m+m'-k]q^{-k-s}+\rho(m',m'-k-1)q^{m+m'-k-1+s}[k+1]\\
&\quad\,-[m]q^{-m'-s}\rho(m',m'-k)-q^{m+s}[m']\rho(m'-1,m'-k-1)\nonumber\\[3mm]
D^-(m,m')&=q^{m+s}[m']\alpha_-(m')-\rho(m',1)[m+1]q^{-m'+1-s}\\&\quad\,
 +[m]q^{-m'-s}\rho(m',1) -q^{m+s}[m']\alpha_-(m'-1)\,.\nonumber
\end{align}
For the creation operator $S_+$ we obtain 
\begin{equation}
\begin{split}
  [\Delta(S_+),\mathcal{H}]|m\rangle\otimes|m'\rangle&=-\sum_{k=0}^{m-1}A_k^+(m,m')|k\rangle\otimes|m'+m-k+1\rangle+B^+(m,m')|m+1\rangle\otimes|m'\rangle\\
 &\quad\,-\sum_{k=0}^{m'-1}C_k^+(m,m')|m+m'-k+1\rangle\otimes|k\rangle+D^+(m,m')|m\rangle\otimes|m'+1\rangle
  \end{split}
\end{equation} 
where $A^+$, $B^+$, $C^+$ and $D^+$  are given by 
% {\small
\begin{align}
A_k^{+}(m,m')&=\rho(m,m-k+1)[k-1+2s]q^{-m'-m+k-1-s}+\rho(m,m-k)q^{k+s}[m'+m-k+2s]\nonumber \\
&\quad\,-[m+2s]q^{-m'-s}\rho(m+1,m-k+1)-q^{m+s}[m'+2s]\rho(m,m-k)\\[3mm]
B^+(m,m')&=[m+2s]q^{-m'-s}\alpha_+(m)-\rho(m',1)q^{m+1+s}[m'+2s-1]\\
 &\quad\,-[m+2s]q^{-m'-s}\alpha_+(m+1)+q^{m+s}[m'+2s]\rho(m'+1,1)\nonumber\\[3mm]
 C_k^+(m,m')&=\rho(m',m'-k)[m+m'-k+2s]q^{-k-s}+\rho(m',m'-k+1)q^{m+m'-k+1+s}[k-1+2s]\nonumber\\
 &\quad\,-[m+2s]q^{-m'-s}\rho(m',m'-k)-q^{m+s}[m'+2s]\rho(m'+1,m'-k+1)\\[3mm]
 D^+(m,m')&=q^{m+s}[m'+2s]\alpha_-(m')-\rho(m,1)[m+2s-1]q^{-m'-1-s}\\
 &\quad\,
 -q^{m+s}[m'+2s]\alpha_-(m'+1)+[m+2s]q^{-m'-s}\rho(m+1,1)\,.\nonumber
\end{align} 
% 
% }
All coefficients $A^\pm$, $B^\pm$, $C^\pm$ and $D^\pm$ vanish. This can be shown using 
\begin{equation}
 \alpha_-(m+1)-\alpha_-(m)=\frac{1}{q^{4 s}-q^{2 m+8 s}}\,,\qquad 
 \alpha_+(m)-\alpha_+(m+1)=\frac{1}{q^{4 s}-q^{-2 m}}\,,
\end{equation} 
which follow from the definition of $\alpha_\pm$ in \eqref{eq:alpha}, and the relations
\begin{align}
 \rho(m,k)&=\frac{\left(q^{2 m}-1\right) \left(q^{2 (m+2 s)}-q^{2 k+2}\right)}{\left(q^{2 m}-q^{2 k}\right) \left(q^{2 (m+2 s)}-q^2\right)}\rho(m-1,k)\,,\\[3mm]
 \rho(m,k)&=\frac{\left(q^{2 k}-q^2\right) q^{2 s-2} \left(q^{2 k}-q^{2 m+2}\right)}{\left(q^{2 k}-1\right) \left(q^{2 k}-q^{2 (m+2 s)}\right)}\rho(m,k-1)\,,
\end{align} 
that arise from the formula for $\rho$ in \eqref{eq:rho}. 

We thus found that the Hamiltonian density defined in \eqref{eq:haction} is $U_q(sl_2)$ invariant. As a consequence the eigenvalues of $\mathcal{H}$ are degenerate and can be obtained by only considering the lowest weight states of the representations of the irreducible tensor product decomposition in \eqref{eq:irrep}. In the next section we verify that the corresponding lowest weights states are eigenstates of the Hamiltonian density and compute the eigenvalue.

\subsection{Eigenvalues of the Hamiltonian density on the lowest weight state}
To identify the Hamiltonian density in \eqref{eq:haction} with the one given in \eqref{ham:bytsko} it remains to compare their action on the lowest weight states of the irreducible tensor product decomposition in \eqref{eq:irrep}.

The lowest weight states can be determined from the condition
\eqref{eq:lws}
by making the ansatz
\begin{equation}\label{eq:state}
 \lws=\sum_{k=0}^jc_{j,k}|k\rangle\otimes|j-k\rangle\,.
\end{equation} 
The conditions in \eqref{eq:lws} then yield the difference equation for the coefficients in \eqref{eq:state}. It reads
\begin{equation}\label{eq:diff}
 [k+1]q^{-s-j+k+1}c_{j,k+1}+[j-k]q^{k+s}c_{j,k}=0\,.
\end{equation} 
Thus up to a normalisation we take
\begin{equation}
c_{j,k}=q^{2k(j+s)}\frac{(q^{2-2j};q^2)_{k-1}}{(q^{4};q^2)_{k-1}}\,
\end{equation}
which can be shown to satisfy \eqref{eq:diff}.

Now, using \eqref{eq:haction}, we compute the action of the Hamiltonian density on the lowest weight states. First we note that after exchanging the sums, that appear in the lowest weight state and in the definition of the Hamiltonian density \eqref{eq:haction}, the action of the latter can be written as
\begin{equation}
\begin{split}
\mathcal{H}\lws&=\sum_{k=0}^j \mathcal{C}(j,k)|k\rangle\otimes|j-k\rangle
 \end{split}
\end{equation} 
where the coefficients read
\begin{equation}
  \mathcal{C}(j,k)=
 c_{j,k}\left(\alpha_+(k)+\alpha_-(j-k)\right)-\sum_{l=k+1}^{j}c_{j,l}\,\rho(l,l-k)-\sum_{l=0}^{k-1}c_{j,l}\,\rho(j-l,k-l)\,.
\end{equation} 
It thus remains to show that 
\begin{equation}\label{eq:ev}
  \mathcal{C}(j,k)=\lambda_jc_{j,k}\,,
\end{equation} 
where 
\begin{equation}
 \lambda_j=\alpha_-(j)+\alpha_+(j)=\frac{\psib(j+2 s)-\psib(2 s)}{-q^{4 s} \log(q)}\,,
\end{equation} 
cf.~\eqref{ham:bytsko}.
This can be done as follows. First we note that
\begin{equation}
\begin{split}
\frac{\mathcal{C}(j,k)}{c_{j,k}}=
\alpha_+(k)+\alpha_-(j-k)&-\sum_{l=1}^{j-k}\frac{q^{2 l (j+s)} \left(q^{2 k-2 j};q^2\right)_l}{\left(q^{2 k+2};q^2\right)_l}\,\rho(l+k,l)
\\&-\sum_{l=1}^{k}\frac{q^{-2 l (s-1)} \left(q^{-2 k};q^2\right){}_l}{\left(q^{2 j-2 k+2};q^2\right){}_l}\,\rho(j-k+l,l)\,,
\end{split}
\end{equation} 
after shifting the boundaries of the sum and using the relations
\begin{equation}
 \frac{c_{j,k+n}}{c_{j,k}}=\frac{q^{2 n (j+s)} \left(q^{2 k-2 j};q^2\right)_n}{\left(q^{2 k+2};q^2\right)_n}\,,\qquad 
 \frac{c_{j,k-n}}{c_{j,k}}=\frac{q^{-2 n (s-1)} \left(q^{-2 k};q^2\right){}_n}{\left(q^{2 j-2 k+2};q^2\right){}_n}\,.
\end{equation} 
As a final step, the relation \eqref{eq:ev} then arises from the identities
\begin{equation}\label{eq:lemma1}
 \alpha_+(j)-\alpha_+(k)=-\sum_{l=1}^{j-k}\frac{q^{2 l (j+s)} \left(q^{2 k-2 j};q^2\right)_l}{\left(q^{2 k+2};q^2\right)_l}\,\rho(l+k,l)
\end{equation} 
and
\begin{equation}\label{eq:lemma2}
 \alpha_-(j)-\alpha_-(j-k)=-\sum_{l=1}^{k}\frac{q^{-2 l (s-1)} \left(q^{-2 k};q^2\right)_l}{\left(q^{2 j-2 k+2};q^2\right)_l}\,\rho(j-k+l,l)\,.
\end{equation} 
Both of them are shown in the following by taking a little journey through the land of special functions, see Section~\ref{sec:proof1} and \ref{sec:proof2} respectively. 

\subsubsection{Proof of relation \eqref{eq:lemma1}}\label{sec:proof1}
In order to show \eqref{eq:lemma1} we define 
\begin{equation}
 \mathcal{F}(j,k)=\sum_{l=1}^{j-k}\frac{q^{2 l (j+s)} \left(q^{2 k-2 j};q^2\right)_l}{\left(q^{2 k+2};q^2\right)_l}\,\rho(l+k,l)\,.
\end{equation} 
As $2k-2j\leq 0$ we can extend the sum to infinity such that
\begin{equation}
\begin{split}
 \mathcal{F}(j,k)&=\frac{q^{2j}(q^{4s} ;q^2)_{k}}{(q^2;q^2)_{k} }
\sum_{l=0}^{\infty}\frac{q^{2 l (j+2s)} }{1-q^{2(l+1)}}\frac{\left(q^{2 k-2 j};q^2\right)_{l+1}}{\left(q^{2 k+2};q^2\right)_{l+1}}\frac{(q^2;q^2)_{l+k+1} }{ (q^{4s} ;q^2)_{l+k+1}}\\
&=q^{2j}\frac{1-q^{2(k-j)}}{1-q^{2(k+2s)}}
\sum_{l=0}^{\infty}\frac{q^{2 l (j+2s)} }{1-q^{2(l+1)}}\frac{\left(q^{2 k-2 j+2};q^2\right)_{l}}{\left(q^{2 k+2+4s};q^2\right)_{l}}\,.
\end{split}
\end{equation}
Here we used $(a,q^2)_{n+k}=(a,q^2)_n(aq^{2n},q^2)_k$ in the second step. Next we write $\mathcal{F}$ as a basic hypergeometric function
\begin{equation}
\begin{split}
 \mathcal{F}(j,k)
&=\frac{q^{2j}}{1-q^2}\frac{1-q^{2(k-j)}}{1-q^{2(k+2s)}}
\sum_{l=0}^{\infty}\frac{\left(q^{2 (j+2s)}\right)^l }{(q^2;q^2)_l}\frac{(q^2;q^2)_l(q^2;q^2)_l\left(q^{2 k-2 j+2};q^2\right)_{l}}{(q^4;q^2)_l\left(q^{2 k+2+4s};q^2\right)_{l}}\\
&=\frac{q^{2j}}{1-q^2}\frac{1-q^{2(k-j)}}{1-q^{2(k+2s)}}\,_3\Phi_2\left(q^2,q^2,q^{2(1-j+k)};q^4,q^{2+2k+4s};q^2;q^{2j+4s}\right)
\end{split}
\end{equation}
cf.~\eqref{eq:bhf}. Serendipitously, such function was studied before in \cite{KRATTEN}. Here the following formula was given
\begin{equation}\label{eq:hyptopsi}
\begin{split}
 \,_3\Phi_2\left(q^2,q^2,q^{2(a+1)};q^4,q^{2(b+1)};q^2;q^{2(b-a)}\right)
 &=\frac{1-q^{-2b}}{1-q^{-2a}}\frac{1-q^2}{2\log(q)}\Big(\psib({b-a})-\psib(b)-a\log(q)\Big)\,,
 \end{split}
\end{equation} 
see \cite[equation (3.4)]{KRATTEN}.
Taking $a=k-j$ and $b=k+2s$, where $b-a=2s+j$, we obtain
\begin{equation}
\begin{split}
 \mathcal{F}(j,k)
&=\frac{1}{2q^{4s}\log(q)}\Big(\psib({2s+j})-\psib(2s+k)+(j-k)\log(q)\Big)=\alpha_+(k)-\alpha_+(j)\,.
\end{split}
\end{equation}
This proves \eqref{eq:lemma1}.
\subsubsection{Proof of relation \eqref{eq:lemma2}}\label{sec:proof2}
To show \eqref{eq:lemma2} we define
\begin{equation}
 {\mathcal{G}}(j,k)=\sum_{l=1}^{k}\frac{q^{-2 l (s-1)} \left(q^{-2 k};q^2\right){}_l}{\left(q^{2 j-2 k+2};q^2\right){}_l}\,\rho(j-k+l,l)\,.
\end{equation} 
Again we can extend the sum to infinity as $-2k\leq0$ and write 
\begin{equation}
\begin{split}
 {\mathcal{G}}(j,k)&=\frac{(q^{4s} ;q^2)_{j-k}}{ q^{4s-2}(q^2;q^2)_{j-k} }\sum_{l=0}^{\infty}\frac{q^{2l } }{1-q^{2(l+1)}}\frac{\left(q^{-2 k};q^2\right)_{l+1}}{\left(q^{2 j-2 k+2};q^2\right)_{l+1}}\,\frac{(q^2;q^2)_{j-k+l+1} }{(q^{4s} ;q^2)_{j-k+l+1}}
 \\&=q^{2-4s}\frac{1-q^{-2k}}{ 1-q^{4s+2(j-k)} }\sum_{l=0}^{\infty}\frac{q^{2l } }{ 1-q^{2(l+1)}}\frac{\left(q^{-2 k+2};q^2\right)_{l}}{(q^{2(2s+j-k+1)} ;q^2)_{l}}\,.
 \end{split}
\end{equation} 
Then as before we write the sum as a basic hypergeometric function
\begin{equation}
\begin{split}
 \mathcal{G}(j,k)
&=\frac{q^{2-4s}}{1-q^2}\frac{1-q^{-2k}}{ 1-q^{4s+2(j-k)} }
\sum_{l=0}^{\infty}\frac{\left(q^{2}\right)^l }{(q^2;q^2)_l}\frac{(q^2;q^2)_l(q^2;q^2)_l\left(q^{-2 k+2};q^2\right)_{l}}{(q^4;q^2)_l(q^{2(2s+j-k+1)} ;q^2)_{l}}\\
&=\frac{q^{2-4s}}{1-q^2}\frac{1-q^{-2k}}{ 1-q^{4s+2(j-k)} }\,_3\Phi_2\left(q^2,q^2,q^{2(1-k)};q^4,q^{2(2s+j-k+1)};q^2;q^{2}\right)\,.
\end{split}
\end{equation}
This is not yet of the form to apply \eqref{eq:hyptopsi}. However, the latter basic hypergeometric function can be written as 
\begin{equation}
\begin{split}
\,_3\Phi_2\big(q^2,q^2,q^{2(1-k)};q^4,q^{2(2s+j-k+1)}&;q^2;q^{2}\big)=q^{2(k-1)}\frac{\left(q^{2(2s+j-k)};q\right)_{k-1}}{\left(q^{2(2s+j-k+1)};q\right)_{k-1}}
\\&\times \,_3\Phi_2\left(q^2,q^2,q^{2(1-k)};q^4,q^{2(2-2s-j)};q^2;q^{2(1-2s-j+k)}\right)\,,
\end{split}
\end{equation} 
see \cite[equation  (III.12)]{gasper2004basic}.
Now we can use \eqref{eq:hyptopsi} again and arrive at
\begin{equation}
 \mathcal{G}(j,k)=\frac{1}{2q^{4s}\log(q)}\Big(\psib(1-2s-j)-\psib(1-2s-j+k)-k\log(q)\Big)\,.
\end{equation} 
This is not quite the result we expected but one may hope for the existence of the q-analog of the reflection formula of the psi-function such that
\begin{equation}
 \psib(1-2s-j)-\psib(1-2s-j+k)=\psib(2s+j)-\psib(2s+j-k)\,.
\end{equation} 
This equation can be derived using the relation 
\begin{equation}
 \frac{\Gamma_{q}(\frac{1}{2})\Gamma_{q}(\frac{1}{2})}{\Gamma_{q}(\frac{1}{2}+x)\Gamma_{q}(\frac{1}{2}-x)}=\cos(\pi x)\sum_{n=1}^\infty\left(\frac{1+2r^{2n}\cos(2\pi x)+r^{4n}}{1+2r^{2n}+r^{4n}}\right)\,,
\end{equation} 
where $\log(q)\log(r)=\pi^2$, see \cite[equation (5.24)]{Askey}. As a consequence we find that 
\begin{equation}
(-1)^k\Gamma_{q}\left(\frac{1}{2}+x\right)\Gamma_{q}\left(\frac{1}{2}-x\right)=\Gamma_{q}\left(\frac{1}{2}+x+k\right)\Gamma_{q}\left(\frac{1}{2}-x-k\right)\,,
\end{equation} 
with $k\in\mathbb{Z}$.
Finally, taking the logarithmic derivative yields
\begin{equation}\label{psi:refl}
 \psib\left(\frac{1}{2}+x\right)-\psib\left(\frac{1}{2}-x\right)=\psib\left(\frac{1}{2}+x+k\right)-\psib\left(\frac{1}{2}-x-k\right)\,.
\end{equation} 
We then find that
\begin{equation}
 \mathcal{G}(j,k)=\frac{1}{2q^{4s}\log(q)}\Big(\psib(2s+j)-\psib(2s+j-k)-k\log(q)\Big)=\alpha_-(j-k)-\alpha_-(j)
\end{equation} 
which concludes the proof of \eqref{eq:lemma2}.

\section{Limiting cases}\label{sec:limit}
\begin{figure}
  \begin{center}  
 \includegraphics[width=0.8\columnwidth]{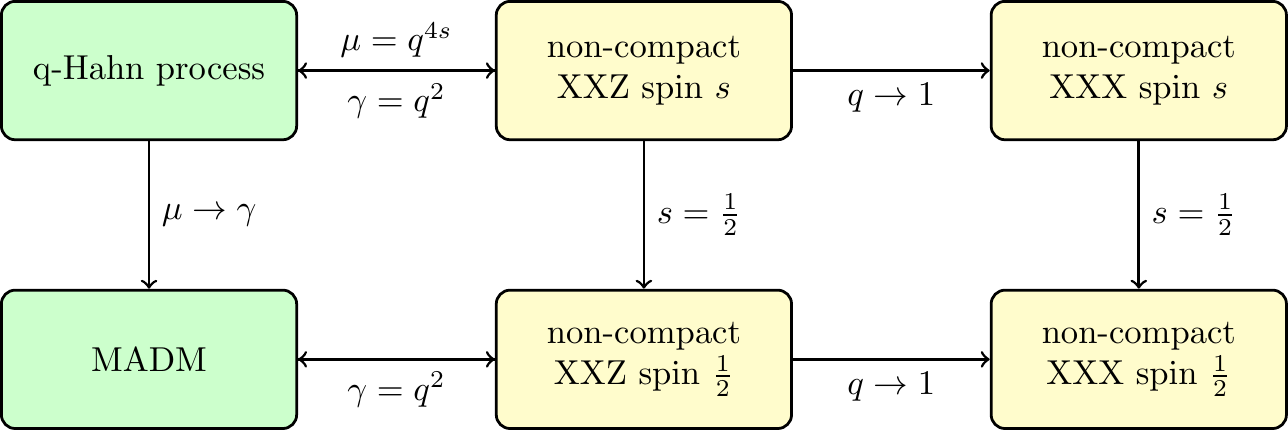}
  \end{center}  
  \caption{Stochastic hopping models in green boxes and corresponding spin chains in yellow boxes. The q-Hahn process reduces to the multi particle asymmetric diffusion model (MADM)   for $\mu\to\gamma$. The identification with the non-compact spin chains is realised by taking $\mu=q^{4s}$ and $\gamma=q^2$. Here $s$ denotes the spin label and the deformation parameter $q$ is related to the anisotropy in the spin chain Hamiltonian. Sending the latter to $q\to1$ we recover the rational limit of the trigonometric spin chains. }
    \label{fig:graph}
\end{figure}
The Hamiltonian of the non-compact XXZ spin chain depends on the parameters $q$ and $s$. In the rational limit $q\to 1$ we recover the non-compact XXX chain, cf.~\cite{Frassek19}. The similarity transformation relating the spin chain and the stochastic process becomes trivial in this limit. The process corresponding to the case $s=\frac{1}{2}$ with parameter $q$ has been studied in \cite{Sasamoto} and is known as the MADM. Its relation in the rational limit with the non-compact Heisenberg chain \cite{Beisertt} has only recently been pointed out in \cite{Frassek19}. See also  Figure~\ref{fig:graph} where the relations between the different models are summarised.

\paragraph{Spin $\frac{1}{2}$ and MADM}
For $s=\frac{1}{2}$ the Pochhammer symbols in the hopping rates disappear and so does the dependence on the number of particles at the initial site. For the spin chain Hamiltonian density one finds 
\begin{equation}
 \rho_{s=\frac{1}{2}}(m,k)=-\frac{1}{q^2(q^{k}-q^{-k})}=-\frac{q-q^{-1}}{q^2}\frac{1}{[k]}
\end{equation} 
and
\begin{equation}
 \alpha_\pm(m)
 =\frac{\psib\left(m+1\right)-\psib\left(1\right)\pm m\log(q)}{-2q^{2}   \log(q)}\,,
\end{equation} 
For the stochastic process setting $s=\frac{1}{2}$ is equivalent to $\mu\to\gamma$. Here we obtain the rates
\begin{equation}
\beta_+(m,k)=\frac{\gamma ^{k} }{\gamma \left(1-\gamma^k\right)}\,,\qquad 
\beta_-(m,k)= \frac{1}{\gamma  \left(1-\gamma^k\right)}\,.
\end{equation} 
They can be identified with the hopping rates of the MADM.

\paragraph{Totally asymmetric limit}
Like in the finite dimensional ASEP the Markov generator reduces to the one of a left or right moving process for $\gamma\to0$ and $\gamma\to\infty$ respectively. Namely for $\gamma\to0$ we find that 
\begin{equation}
 \lim_{\gamma\to0}\gamma^{2s} \beta_+(m,k)|_{\mu=\gamma^{2s}}=0\,,\qquad   \lim_{\gamma\to0}\gamma^{2s} \beta_-(m,k)|_{\mu=\gamma^{2s}}=1\,,
\end{equation} 
and
\begin{equation}
 \lim_{\gamma\to0}\gamma^{2s} \alpha_+(m)|_{\mu=\gamma^{2s}}=0\,,\qquad   \lim_{\gamma\to0}\gamma^{2s} \alpha_-(m)|_{\mu=\gamma^{2s}}=m\,,
\end{equation} 
for $2s\in \mathbb{N}$. While for the case $\gamma\to\infty$ we get 
\begin{equation}
 \lim_{\gamma\to\infty}\gamma^{2s} \beta_+(m,k)|_{\mu=\gamma^{2s}}=1\,,\qquad   \lim_{\gamma\to\infty}\gamma^{2s} \beta_-(m,k)|_{\mu=\gamma^{2s}}=0\,,
\end{equation} 
while 
\begin{equation}
 \lim_{\gamma\to\infty}\gamma^{2s} \alpha_+(m)|_{\mu=\gamma^{2s}}=m\,,\qquad   \lim_{\gamma\to\infty}\gamma^{2s} \alpha_-(m)|_{\mu=\gamma^{2s}}=0\,.
\end{equation}

\paragraph{Rational limit} In the rational limit we recover the non-compact XXX spin chain. The hopping rates are of the form
\begin{equation}
 \lim_{q\to 1}\log(q^{-1})\rho(m,k)=\frac{1}{2k}\frac{\Gamma (m+1) \Gamma (m-k+2s)}{\Gamma (m-k+1) \Gamma (m+2 s)}
\end{equation} 
and
\begin{equation}
 \lim_{q\to 1}\log(q^{-1})\alpha_\pm(m)
 =\frac{\psi\left(m+2s\right)-\psi\left(2s\right)}{2}\,,
\end{equation} 
We stress again that the similarity transformation which relates the local Markov generator to the Hamiltonian density becomes trivial in this case. 

\paragraph{TAZRP}
Taking a limit that can be related to a single totally asymmetric zero range process (TAZRP), cf.~\cite{Kuniba,Kuniba:2016tzg}, does not seem to be straightforward at the level of the Hamiltonian density. However, as a consequence of Baxter's TQ-equation which relates the transfer matrix and the Q-operator, the generators of the two TAZRP's may arise from the logarithmic derivative of the Q-operator of the non-compact XXZ chain at two special points of the spectral parameter.
Such mechanism was demonstrated for the non-compact spin $\frac{1}{2}$ XXX chain in \cite[Appendix C]{Frassek:2012mg} using an oscillator construction of Q-operators going back to \cite{Bazhanov:1998dq,Bazhanov:2010ts,Frassek:2011aa}.

\section{Conclusion}\label{sec:conc}
In this note we gave an explicit expression for the action of the Hamiltonian density of the spin $s$ non-compact XXZ chain on the tensor product of two sites of the spin chain \eqref{eq:haction}. Further we showed that it directly relates to the local generator of a continuous-time Markov process \eqref{eq:locmarkov} via \eqref{eq:hammag}, cf.~\eqref{sim:trans} for the whole chain. The rates of the particles hopping are identified with a q-Hahn process studied previously in  \cite{barraquand,Sasamoto,Povolotsky}. 

The identification of the integrable spin chain and the stochastic particle process allows to describe the system in the standard framework of the quantum inverse scattering method. The latter immediately provides a huge variety of integrability methods,  like the algebraic Bethe ansatz, separation of variables and functional methods, to study the model and its limits. Here in particular it might be instructive to study the large spin limit which may be related to the q-Boson totally asymmetric process  \cite{Sasamoto_1998} whose Lax matrices are known to arise in that limiting case. Having understood the algebraic structure of the q-Hahn process for the bulk may further allow to derive the appropriate stochastic boundary conditions from the boundary Yang-Baxter equation as done for the rational limit in \cite{Frassek19}. Further, one may expect that duality and the limit of fluctuating hydrodynamics which were studied in the previous reference do have a q-analog. It would be natural to study whether the Hamiltonian density of non-compact spin chains with higher rank allows a similar identification, see e.g. \cite{Frassek:2011aa} where the analogs of \eqref{ham:bytsko} were discussed in the  rational limit. In particular, we expect that the Hamiltonian of  spin chains with non-compact symmetric representations of higher rank can be related to the multi-species particle process in \cite{Kuniba:2016tzg,Kuniba} with the identifications discussed in at the end of Section~\ref{sec:markov}.

Our findings suggest that the local charges which arise from the Q-operator of the non-compact XXZ spin chain are directly connected to the Markov generator of the totally asymmetric zero range process (TAZRP) which arises from the transfer matrix constructed from the stochastic R-matrix for $U_q(A_1^{(1)})$.
It would be interesting to study this relation in detail.
The oscillator type construction of Q-operators, mentioned in Section~\ref{sec:limit}, has been carried out for non-compact rational spin chains in \cite{Frassek:2011aa,Frassek:2017bfz}, their trigonometric counterpart was so far only studied for the fundamental representation, cf.~\cite{Bazhanov:2008yc,Boos:2012tf}, and deserves further investigation. We refer the reader to  \cite{Chicherin:2012jz,Mangazeev:2014bqa} and references therein for alternative approaches.

\section*{Acknowledgement}
I like to thank Cristian Giardin\`{a},  Mikhail Isachenkov and Jorge Kurchan for useful discussions and interest in the problem. 
R.F. was supported by the {\small IH\'{E}S} visitor program.

\appendix
\section{Special functions}\label{app:qan}
In this appendix we collect some definitions of the q-analogs of some  special functions used in the main text. 
As mentioned before we were using the conventions of \cite{Bytsko:2001uh} such that 
\begin{equation}
 \gammab(x)=\left(q^{-1}-q\right)^{1-x} q^{-\frac{1}{2} (x-1) x}\frac{(q^2;q^2)_\infty}{\left(q^{2x};q^2\right)_\infty}\,,
\end{equation}  
for $|q|<1$. The q-analog of the $\psi$-function can then be written as
\begin{equation}
  \psib(z)=\frac{\partial}{\partial z}\log \Gamma_q(z)=-\log(1-q^2)+2\log(q)\sum_{k=0}^\infty \frac{q^{2(k+z)}}{1-q^{2(k+z)}}+\frac{1}{2} (3-2 z) \log (q)\,.
\end{equation} 
These conventions differ from the definitions which seem to be commonly used. Alternatively the q-analog of the $\Gamma$-function is defined as
\begin{equation}
 \widetilde \Gamma_\gamma(z)=(1-\gamma)^{1-z}\frac{(\gamma;\gamma)_\infty}{\left(\gamma^z;\gamma\right)_\infty}\,,
\end{equation} 
for $|\gamma|<1$. The $\psi$-function then reads
\begin{equation}
  \psim_\gamma(z)=\frac{\partial}{\partial z}\log  \widetilde\Gamma_\gamma(z)=-\log(1-\gamma)+\log(\gamma)\sum_{k=0}^\infty \frac{\gamma^{k+z}}{1-\gamma^{k+z}}\,.
\end{equation} 
The two conventions are related via
\begin{equation}
  \Gamma_{q}(x)=q^{-\frac{1}{2} (x-2) (x-1)}  \widetilde\Gamma_{q^2}(x)\,,
\end{equation} 
and
\begin{equation}
\psib(x)=\psim_{q^2}(x)+\frac{1}{2} (3-2 x) \log (q)\,.
\end{equation} 
Finally we give the definition of the basic hypergeometric function
\begin{equation}\label{eq:bhf}
 \,_3\Phi_2\left(a,b,c;d,e;q;z\right)=
\sum_{l=0}^{\infty}\frac{(a;q)_l(b;q)_l(c;q)_{l}}{(d;q)_l\left(e;q\right)_{l}}\frac{z^l }{(q;q)_l}\,.
\end{equation}

{
%   \small
\bibliographystyle{utphys2}
\bibliography{refs}
  }
  
\noindent\rule{6cm}{0.4pt}

\texttt{Contact: rfrassek@mpim-bonn.mpg.de}

\end{document}